\documentclass[sigconf, nonacm, pdfa]{acmart}
\usepackage{svg}
\newcommand\vldbdoi{XX.XX/XXX.XX}
\newcommand\vldbpages{XXX-XXX}
\newcommand\vldbvolume{19}
\newcommand\vldbissue{12}
\newcommand\vldbyear{2026}
\newcommand\vldbauthors{\authors}
\newcommand\vldbtitle{\shorttitle} 
\newcommand\vldbavailabilityurl{https://github.com/ruc-datalab/DeepAnalyze/tree/main/demo/chat_v2}
\newcommand\vldbpagestyle{empty}

\begin{document}
\title{DA-Studio: An Agentic System for End-to-End Data Analysis}

\author{Yizhe Liu}
\affiliation{%
  \institution{Renmin University of China}
}
\email{liuyizhe2004@126.com}

\author{Shaolei Zhang}
\affiliation{%
  \institution{Renmin University of China}
}
\email{zhangshaolei98@ruc.edu.cn}

\author{Ju Fan}
\affiliation{%
  \institution{Renmin University of China}
}
\email{fanj@ruc.edu.cn}


\begin{abstract}
Real-world data analysis is a multi-step process over heterogeneous inputs rather than merely producing a final answer. A practical system should autonomously organize multi-step workflows, execute generated code in a sandboxed and controllable environment, and remain inspectable through visible action traces and intermediate artifacts. Existing LLM-based analysis tools, however, often emphasize isolated subtasks, leaving limited support for complete execution-grounded workflows.

We present \textbf{DA-Studio (Data Analysis Studio)}, an interactive web-based demo system for end-to-end data analysis that is autonomous, sandboxed, and inspectable. DA-Studio integrates an action-structured analysis backend, a sandboxed execution workspace, and a browser interface for task setup, streamed action traces, artifact preview, code editing and rerunning, and report export. Through iterative action generation, code execution, and feedback incorporation, it incrementally constructs executable analysis steps from raw files and natural-language requests while exposing intermediate results and artifacts throughout the process.
\end{abstract}

\maketitle

\pagestyle{\vldbpagestyle}
\begingroup\small\noindent\raggedright\textbf{PVLDB Reference Format:}\\
\vldbauthors. \vldbtitle. PVLDB, \vldbvolume(\vldbissue): \vldbpages, \vldbyear.\\
\href{https://doi.org/\vldbdoi}{doi:\vldbdoi}
\endgroup
\begingroup
\renewcommand\thefootnote{}\footnote{\noindent
This work is licensed under the Creative Commons BY-NC-ND 4.0 International License. Visit \url{https://creativecommons.org/licenses/by-nc-nd/4.0/} to view a copy of this license. For any use beyond those covered by this license, obtain permission by emailing \href{mailto:info@vldb.org}{info@vldb.org}. Copyright is held by the owner/author(s). Publication rights licensed to the VLDB Endowment. \\
\raggedright Proceedings of the VLDB Endowment, Vol. \vldbvolume, No. \vldbissue\ %
ISSN 2150-8097. \\
\href{https://doi.org/\vldbdoi}{doi:\vldbdoi} \\
}\addtocounter{footnote}{-1}\endgroup

\ifdefempty{\vldbavailabilityurl}{}{
\vspace{.3cm}
\begingroup\small\noindent\raggedright\textbf{PVLDB Artifact Availability:}\\
The source code, data, and/or other artifacts have been made available at \url{\vldbavailabilityurl}.
\endgroup
}

\section{Introduction}

Domain experts often need to analyze heterogeneous data, yet turning raw files into executable analysis steps and final deliverables still requires substantial manual effort. Recent LLM-based systems such as LIDA, Data Formulator 2, and nvAgent have improved natural-language-driven visualization authoring and exploratory analysis~\cite{lida,dataformulator2,nvagent}. However, these systems mainly focus on visualization authoring or other isolated subtasks, leaving users to bridge the gap from raw inputs to executable analysis steps and final deliverables.

\begin{figure}[t]
  \centering
  \includegraphics[width=\linewidth]{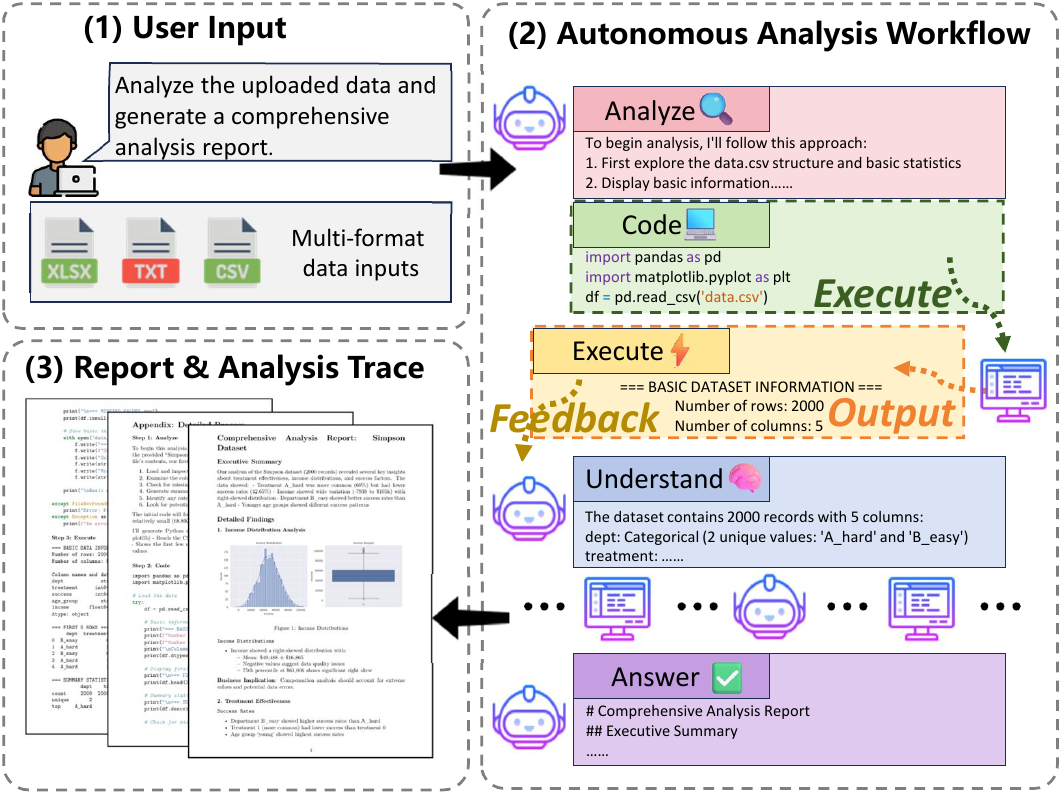}

  \caption{Overview of DA-Studio. Given user data and a natural-language analysis task, DA-Studio orchestrates structured analysis actions, executes generated code in a sandboxed workspace, incorporates execution feedback, and exposes the evolving analysis trace and artifacts to the user.}
  \Description{A workflow diagram of DA-Studio. A user provides data files and a natural-language task, the backend alternates between structured analysis actions, code generation, sandboxed execution, and execution feedback, and the interface exposes traces, artifacts, and final report export.}
  \label{fig:overview}
\end{figure}

For practical end-to-end data analysis, a system should meet three requirements. First, it should autonomously orchestrate multi-step workflows over heterogeneous inputs without requiring users to manually specify a pipeline. Second, it should execute generated code in a sandboxed and controllable environment over user data. Third, it should make the analysis process inspectable through visible action traces and intermediate artifacts.

Recent execution-grounded LLM systems have shown that models can iteratively generate code, execute it, and refine subsequent actions based on execution results~\cite{pal,taskweaver}. To bring these capabilities into an end-to-end data analysis setting, we present \textbf{DA-Studio (Data Analysis Studio)}, an interactive web-based demo system for autonomous, sandboxed, and inspectable data analysis. DA-Studio combines an action-structured analysis backend, a sandboxed execution environment, and a browser interface for streamed traces, artifact preview, code editing and rerunning, and Markdown/PDF report export. Through this interface, users can inspect not only the final results but also the process that produces them. This end-to-end visibility makes autonomous analysis easier to inspect, revise, and reuse in practice.

\vspace*{0.3em}
\noindent\textbf{Overview.} Figure~\ref{fig:overview} illustrates the end-to-end workflow of DA-Studio. \textbf{(1)} A user first specifies an analysis task in natural language and uploads input files. \textbf{(2)} Given the task and the current session context, the backend iteratively selects the next analysis action, such as planning, data inspection, code generation, result interpretation, or answer generation. When a \texttt{<Code>} action is produced, the corresponding script is executed in the sandboxed environment, and the returned outputs are fed back into the session context for subsequent actions. This loop continues until the system produces the final answer. Throughout the workflow, DA-Studio streams action traces, execution feedback, and generated artifacts to the interface so that users can inspect intermediate results and intervene when needed. \textbf{(3)} After the task is completed, DA-Studio exports the final report together with the generated artifacts for later reuse.

\vspace*{0.3em}
\noindent\textbf{Demonstration Workflow.} In the demonstration, users interact with the system by issuing natural-language analysis requests over heterogeneous inputs such as tables and documents. Rather than showing only a final answer, the interface exposes action traces, executable code, intermediate artifacts, and exported reports, allowing the audience to follow a complete execution-grounded analysis session from task setup to final deliverables. A demonstration video is available on YouTube.\footnote{\url{https://youtu.be/CnKEGfunLAI}}

In summary, DA-Studio makes three contributions. First, it supports autonomous, execution-grounded multi-step analysis through an action-structured backend. Second, it provides sandboxed execution with session-scoped workspaces and artifact management for reproducible analysis and rerunning. Third, it offers an inspectable browser interface that unifies data upload, streamed traces, artifact preview, code revision, and report export.

\begin{figure*}[t]
  \centering
  \includegraphics[width=\linewidth]{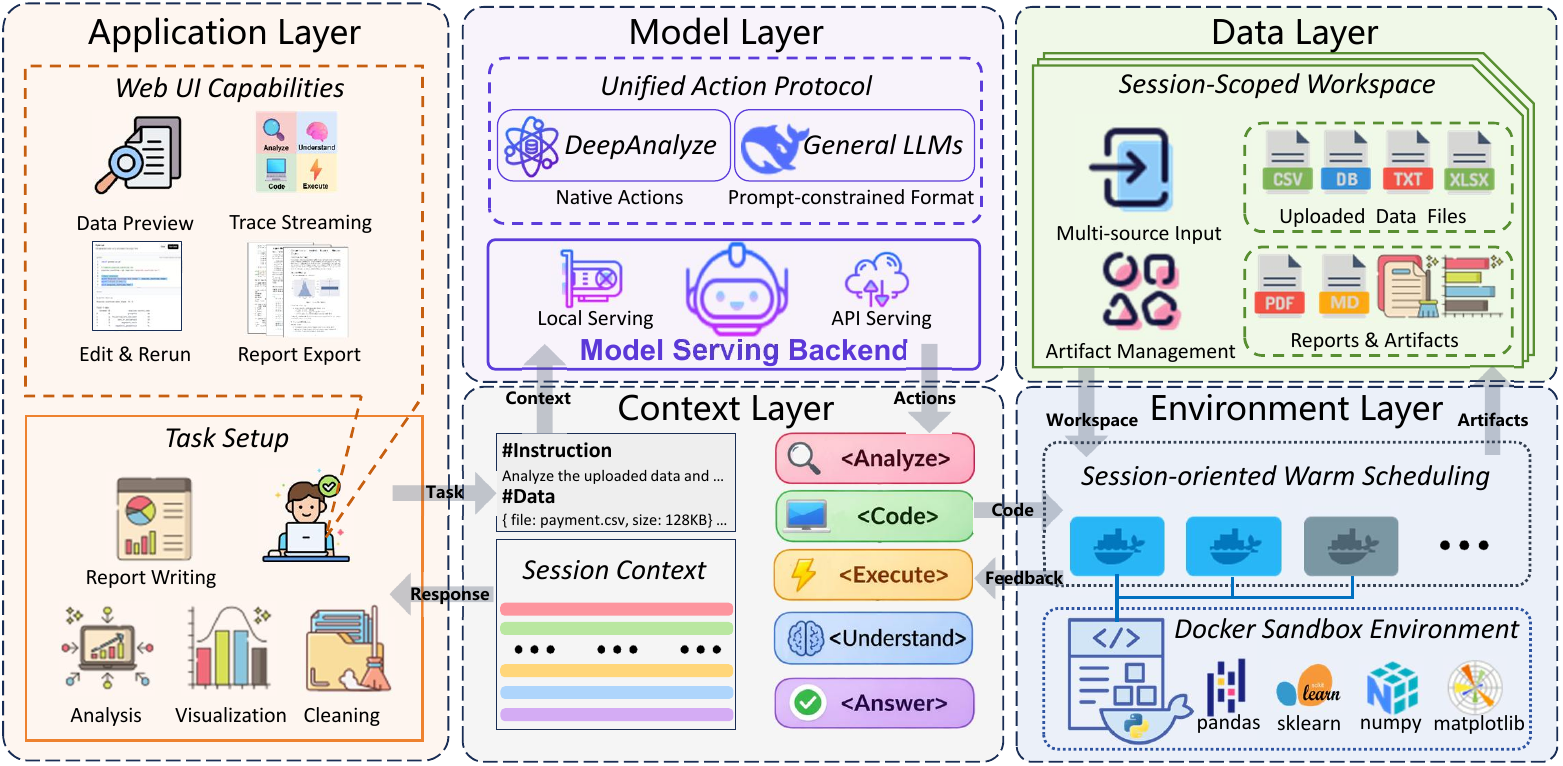}
  \caption{Five-layer architecture with three functional views. The Application Layer supports inspectable interaction, the Model and Context Layers enable autonomous multi-step analysis, and the Data and Environment Layers ensure sandboxed and controllable execution over user data.}
  \Description{A five-layer architecture diagram of DA-Studio. The Application Layer provides user-facing interaction, the Model and Context Layers maintain the iterative analysis loop, the Environment Layer executes generated code in a sandbox, and the Data Layer stores uploaded files and produced artifacts.}
  \label{fig:arch}
\end{figure*}

\section{System Architecture}
To realize autonomous, sandboxed, and inspectable end-to-end data analysis, DA-Studio is organized as a five-layer architecture shown in Figure~\ref{fig:arch}. We describe this architecture through three functional views that expose the system's key technical mechanisms: the Application Layer for inspectable interaction, the Model and Context Layers for autonomous multi-step analysis, and the Data and Environment Layers for sandboxed execution and artifact management.

As illustrated in Figure~\ref{fig:arch}, these layers are linked by the flow of task requests, structured actions, execution feedback, and artifacts across the system. Together, they form a session-scoped workflow from task specification to sandboxed execution, artifact tracking, and user intervention.

\subsection{Application Layer}

To support inspectable and interactive analysis, the Application Layer manages user requests, streamed outputs, and workspace state rather than serving as a passive front end. Each user request is organized as a session-scoped task configuration that bundles the natural-language instruction, selected files, model/runtime parameters, and session identifier. The configuration is forwarded to the lower layers, while uploaded files are immediately registered in the session workspace managed by the Data Layer.

During execution, the Application Layer incrementally renders structured action blocks, code snippets, execution feedback, and final answers, while synchronizing the workspace tree, artifact preview pane, and code editor with the same session state. The interface maintains two synchronized views: the streamed trace shows the current analysis actions and execution feedback, while the workspace view exposes the files and intermediate artifacts materialized in the session. This design allows users to inspect evolving outputs without waiting for the workflow to finish.

The Application Layer also supports post-execution intervention. Generated code blocks can be opened, edited, and rerun in the same workspace, and the final report together with the produced artifacts can be exported in Markdown/PDF form. Consequently, the Application Layer serves as both a visible interface for streamed outputs and an interactive entry point for code revision and rerunning.

\subsection{Model and Context Layers}

To support autonomous multi-step analysis without relying on manually predefined workflows, the Model and Context Layers organize model outputs through a unified action protocol and execute them in an iterative generation--execution--feedback process.

\vspace*{0.3em}

\noindent\textbf{Model Layer.} The Model Layer standardizes backend outputs into a unified action protocol for multi-step data analysis. Rather than following a manually predefined workflow, the model emits the next action on demand, choosing among planning, data inspection, code generation, result interpretation, and final answer generation. Following the action format used in DeepAnalyze~\cite{deepanalyze}, DA-Studio adopts tags such as \texttt{<Analyze>}, \texttt{<Understand>}, \texttt{<Code>}, \texttt{<Execute>}, and \texttt{<Answer>} to format model outputs as structured analysis actions, enabling the model to organize these actions and autonomously select the next step.

DA-Studio further adds a model-compatibility layer on top of this interface. When using DeepAnalyze, the backend can directly consume the model's native action blocks; when using general-purpose LLMs, it uses a prompt-constrained compatibility layer to enforce the same output format. In both cases, an output parser validates the structure and completeness of generated action blocks, and code is dispatched for execution only after a complete \texttt{<Code>} segment is finalized. As a result, heterogeneous models can be normalized into the same executable action interface, while code execution remains gated by explicit structural checks.

\vspace*{0.3em}

\noindent\textbf{Context Layer.} The Context Layer maintains the evolving state required by this protocol. Instead of placing all uploaded content directly into the prompt, it follows a data-oriented context construction strategy similar in spirit to DeepAnalyze~\cite{deepanalyze}. The task instruction and lightweight file descriptors are organized into a structured prompt, rather than placing the full contents of input files directly into the context. The model can then retrieve relevant data contents on demand through code execution and the returned results. This avoids treating heterogeneous raw files as static prompt text and lets the model interact with external data sources on demand.

Operationally, the Context Layer maintains a rolling session state consisting of task instructions, prior action traces and execution outputs. When a completed \texttt{<Code>} block is produced, the corresponding script is extracted and dispatched to the Environment Layer. Returned program outputs are wrapped into a \texttt{<Execute>} block and reinjected into the context for the next round. As a result, the backend loop becomes a generation--execution--feedback cycle rather than a one-shot response.

\subsection{Data and Environment Layers}
To ensure sandboxed and controllable execution over user data, the Data and Environment Layers provide isolated runtime support and workspace-level file management.
\vspace*{0.3em}

\noindent\textbf{Data Layer.} The Data Layer manages both uploaded inputs and generated outputs throughout the task through a session-scoped workspace abstraction. Each session is assigned an independent workspace root, and all file operations such as upload, preview, and cleanup are resolved against this root through path-safe APIs. This organization provides file-level isolation for executing the generated code.

Beyond raw storage, the Data Layer also serves as an artifact-management component. Newly generated outputs such as cleaned tables, plots, and summaries are registered in a dedicated generated namespace together with an artifact index, allowing the system to distinguish user-provided inputs from system-created derivatives. This indexing mechanism supports several user-facing capabilities, including incremental browsing of generated outputs during execution, bundled export of analysis results, and artifact references that can be reused when rerunning code and assembling the final report.

\vspace*{0.3em}

\noindent\textbf{Environment Layer.} The Environment Layer executes model-generated code inside a Docker-based sandbox with a stable Python data-analysis stack. Each \texttt{<Code>} block is materialized as a temporary script in the session workspace and executed in a container whose working directory is mounted to that workspace, confining file access to the current session and enabling controllable execution of generated scripts. Containers are reclaimed after inactivity, and each execution remains scoped to the current session workspace.

To reduce latency in iterative analysis, DA-Studio adopts \emph{session-oriented warm scheduling}. The first execution request of a session triggers allocation of a dedicated container, and subsequent requests from the same session reuse the warm container; different sessions are mapped to different containers, and idle containers are reclaimed after timeout. Before and after each execution, the system snapshots the workspace to detect added or modified files, registers them as generated artifacts, and returns corresponding references to upper layers.

\begin{figure*}[t]
  \centering
  \includegraphics[width=\linewidth]{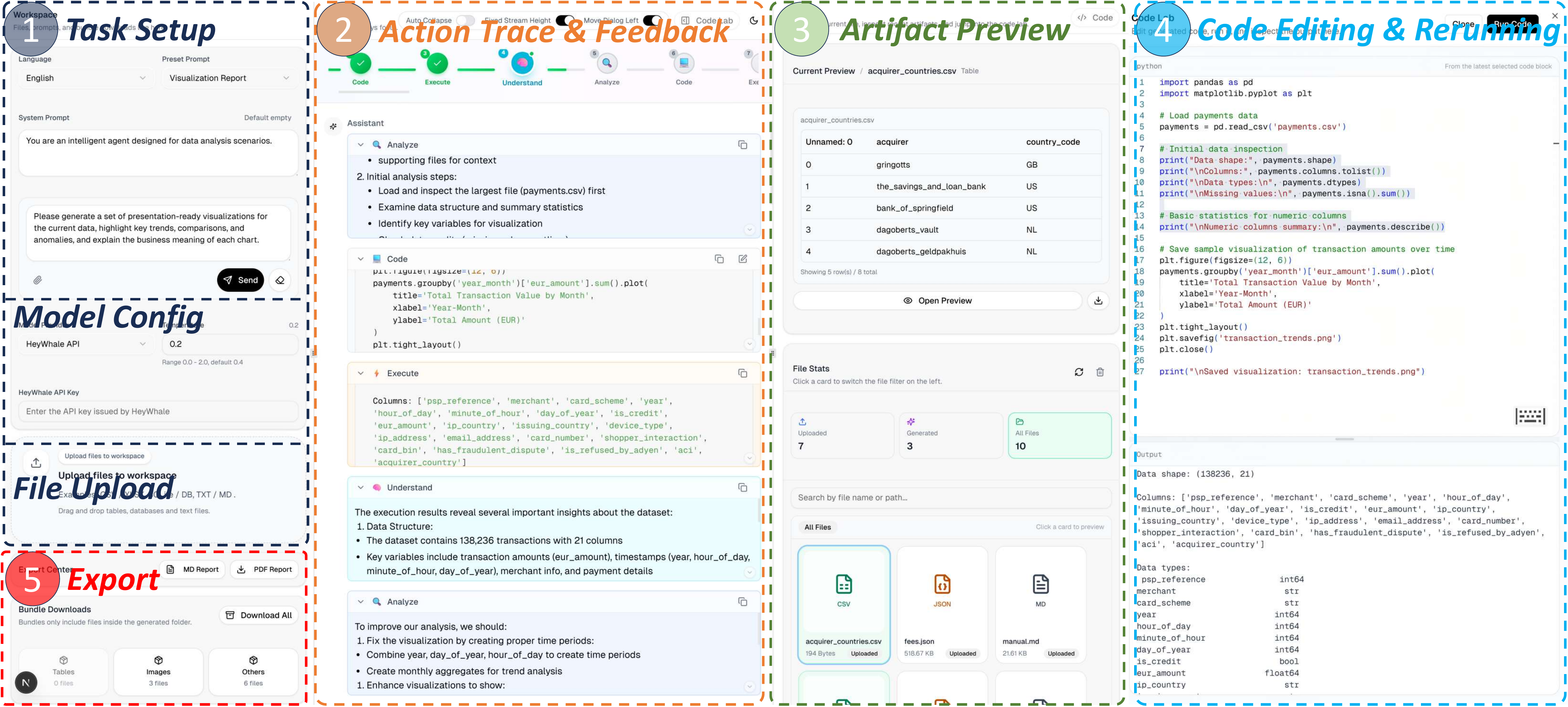}
  
  \caption{Screenshot of DA-Studio during a transaction-analysis demo. Region (1) supports task setup over heterogeneous files, (2) streams action traces and execution feedback, (3) previews intermediate artifacts, (4) enables code editing and rerunning, and (5) exports reports and bundled outputs.}
  \Description{An annotated screenshot of DA-Studio with five highlighted regions: task setup on the left, streaming action traces in the center, artifact preview on the right, a code editor for rerunning, and export controls for reports and generated files.}
  \label{fig:demo}
\end{figure*}


\section{Demonstration Scenarios}

The demonstration shows how DA-Studio supports execution-grounded data analysis over multi-file inputs. Figure~\ref{fig:demo} presents a representative merchant-payment analysis task in which the user uploads a multi-format dataset and issues a natural-language request. The five highlighted regions in Figure~\ref{fig:demo} correspond to the main stages of this workflow.

\noindent\underline{\textit{(1) Starting from task setup.}} As shown in the left panel of Figure~\ref{fig:demo}, the workflow begins with a user uploading a merchant-payment dataset and specifying an analysis request in natural language. In this running example, the workspace contains data and supporting documents in multiple formats, including \texttt{payments.csv}, \texttt{fees.json}, and \texttt{manual.md}. The user then inputs a task instruction or chooses a preset prompt for common tasks such as exploratory analysis or visualization-oriented reporting. The interface additionally exposes model configuration options, including the system prompt, decoding temperature, and model selection.

\noindent\underline{\textit{(2) Following the analysis as it unfolds.}} After submission, the user's attention shifts to the central streaming panel. Here, DA-Studio reveals the ongoing analysis trace step by step, rather than waiting to show only a final answer. In this running example, the model first uses an \texttt{<Analyze>} block to plan the initial steps, and then emits a \texttt{<Code>} block to generate code for reading and processing the relevant files. After receiving execution feedback through an \texttt{<Execute>} block, it enters an \texttt{<Understand>} block to interpret the data structure and contents, and then returns to \texttt{<Analyze>} to plan the next action. This iterative process is continuously streamed in the central panel of Figure~\ref{fig:demo}, allowing users to clearly follow how the analysis progresses from one step to the next.

\noindent\underline{\textit{(3) Inspecting intermediate artifacts.}} While the streamed trace explains what the system is doing, the right-side workspace in Figure~\ref{fig:demo} shows what the system has produced so far. Uploaded files remain visible next to newly generated artifacts such as cleaned tables, figures, and execution summaries. The user can click file cards in the right-side panel to preview the contents of either uploaded inputs or newly generated artifacts in real time. This workspace-level preview makes inspectability concrete by allowing users to examine both supporting files and intermediate outputs without leaving the current session.

\noindent\underline{\textit{(4) Revising and rerunning generated code.}} Once the main interaction loop has produced a result, the user can continue working with the generated scripts through the integrated code editor, which supports convenient adjustment and reuse of existing analysis code. Instead of restarting the full analysis from scratch, the user may open a code block, make targeted changes by providing modification instructions in natural language or by editing the code directly, and rerunning the script in the same sandboxed workspace. 

\noindent\underline{\textit{(5) Exporting final results and analysis traces.}} The workflow ends in the export area highlighted in Figure~\ref{fig:demo}. After reviewing the outputs, the user can export the final report in Markdown or PDF form and download the generated artifacts as a bundled archive. The exported report summarizes both the final findings and the corresponding analysis trace, while the bundled outputs preserve the tables, figures, and other files produced during execution.

Overall, Figure~\ref{fig:demo} shows DA-Studio's end-to-end workflow: users start from raw files and a natural-language goal, follow the execution-grounded analysis, inspect artifacts, revise code, and export outputs. This workflow illustrates DA-Studio as a demo system for autonomous, sandboxed, and inspectable data analysis.

\bibliographystyle{ACM-Reference-Format}
\bibliography{sample}

\end{document}